\documentclass[11pt,twoside]{article}

\usepackage{asp2006}
\usepackage{epsf}
\usepackage{psfig}
\usepackage{lscape}

\begin{document}
\title{The study of resonant variability observed in the massive LMC system BI 108}   
\author{Zbigniew Ko{\l}aczkowski\altaffilmark{1,2}, Ronald Mennickent\altaffilmark{1} \& Thomas Rivinius\altaffilmark{3}  }   
\altaffiltext{1}{Departamento de Astronom\'{\i}a, Universidad de Concepci\'on, Casilla 160-C, Concepci\'on, Chile}
\altaffiltext{2}{Instytut Astronomiczny Uniwersytetu Wroclawskiego, Kopernika 11, Wroclaw, Poland
}
\altaffiltext{3}{ESO Paranal Observatory
}

\begin{abstract} 
The LMC star BI 108 is photometrically variable with the unique light curve: 
Two strong periods are present in a strict 3:2 resonance, staying coherent over 
several observing seasons. A spectroscopic data collected at VLT/UVES reveals 
unexpected and still not fully understood behavior of the system, that has not been observed 
in any other early-type multiple systems. 
\end{abstract}


\section{Photometric properties of the system}

An object LMC SC9-127519 was discovered as a variable star with unique photometric properties
during our multiperiodicity search in the OGLE-II database (Szyma\'nski, 2005).
Independently this object was announced as two binary systems in 3:2 resonance by Ofir (2008).
With the coordinates, the magnitude (V=13.3) and the colour (B-V=-0.142), we identified object 
as BI 108 in the list of Brunet et al. (1975), for which they give a spectral 
type of B1-2. Schmidt-Kaler et al. (1999) list B1:II: for the same object. 
Probably BI 108 is one of the brightest member of a young open cluster, NGC 1881.
  The OGLE-II data spans almost four years. A time series analysis returns two periods.
Of these two periods the stronger one ($P_1$=3.57793 d), with a peak-to-peak amplitude of 
about 0.18 mag, is clearly of a double wave nature. A light curve of such that particular 
appearance can, with little doubt, be attributed to an eclipsing binary nature (Fig.~1). 
The weaker period, with about 0.05 mag peak-to-peak, has the value of $P_2$=5.36654 d. 
Within the uncertainties this is exactly 3/2 of the stronger period. 
The second light curve is not sinusoidal (Fig.~1).
An analysis of the photometric variability allows two options, including $P_2$ or $P_2$/2, 
and 3:2 or 3:4 resonance values, respectively.
With the MACHO data (ID 79.5378.25) we were able to confirm both periodicities with the same 
amplitudes in three bands and their long term coherence.
Moreover both periods have exactly the same moment of the minimum. It is an additional constraint 
difficult to explain in the model. Such a clearly resonant and long-term stable period relation 
is not known for any other early type variable star.

\begin{figure}[!t]
\plotone{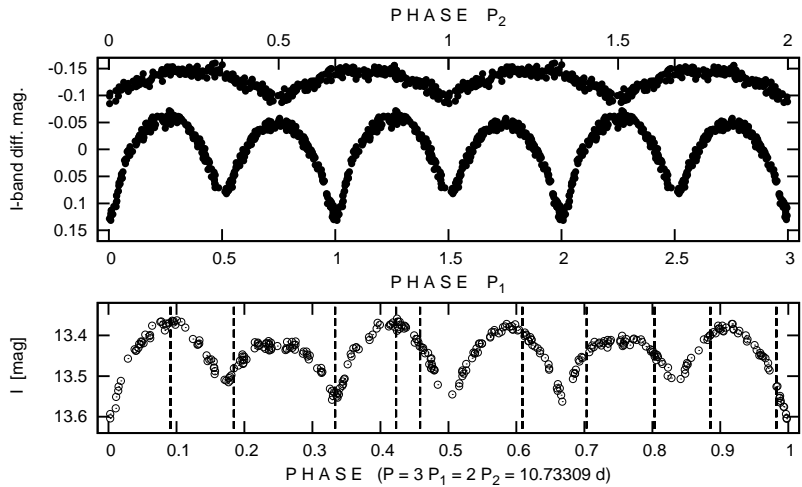}
\caption{ Upper panel: The OGLE I-band light curves for both periods of the BI 108
system. For each period the contribution of the respective other one has been removed.
Lower panel: The original light curve folded with superperiod. In all cases 
$HJD_0$ = 2451163.8915. Dashed lines indicate phases of the spectral observations.  
 }
\end{figure}


\section{Preliminary spectroscopic results}
The immediate goal of our follow-up program was to confirm the assumption of an eclipsing binary 
for $P_1$ period, and clarify the nature of the period with smaller amplitude, 
which could be either $P_2$ in case an ellipsoidal light curve arising from a multiple co-planar 
system, or twice shorter in case of a single wave, e.g.~tidally forced pulsation. 
These processes have their own, distinct spectroscopic signatures revealed in the radial 
velocity and line shape changes. 
  Ten epochs spectroscopic observations were carried out at ESO VLT/UVES. 
 The system has SB2 nature and contains two similar components with spectral features of 
B1 II/Ib stars. But surprisingly an orbital period seems to be $P_2$. 
We do not find modulation of the radial velocities with $P_1$ period and moreover we do not 
see any signature of the third component in the system. This result does not support our predictions. 
Only evidence of the shorter variability is the H and He I profile variability.  
When both phases are close to quadrature the system appears as SB2 with almost identical lines but changes relative depth when the shorter cycle is close to minimum. 
This seems to follow a shorter photometric cycle and strongly suggest a variability of one (less massive) component. It could be an effect of unusual eclipses or rotational modulation. But our few spectral observations do not allow to test these hypotheses, and so far the nature of the phenomena 
observed in the BI 108 remains unknown.


\end{document}